\documentclass[%
aip,%
amsmath,amssymb,
reprint,%
longbibliography,
]{revtex4-2}

\usepackage{graphicx}% Include figure files
\usepackage{dcolumn}% Align table columns on decimal point
\usepackage{bm}% bold math
\usepackage{braket}
\usepackage{epstopdf}

\usepackage{fancyhdr}
\fancypagestyle{firstpage}{%
	\fancyhead{}  % remove the default header 
	\fancyhead[C]{The following article has been submitted to  \textit{Journal of Chemical Physics}. After it is published, it will be found at https://publishing.aip.org/resources/librarians/products/journals/}

	\fancyfoot{} % remove the footer 
}

\begin{document}
    
    %\preprint{}
    \title{Exchange-correlation potential built on the derivative discontinuity of electron density}

    \author{Chen Huang}    
    
    \email{chuang3@fsu.edu}  
      
    \affiliation{Department of Scientific Computing, Materials Science and Engineering Program, and National High Magnetic Field Laboratory, Florida State University, Tallahassee, Florida 32306, USA}
    
    \date{\today}
    
    \begin{abstract}

    Electronic structures are fully determined by the exchange-correlation (XC) potential. In this work, we develop a new method to construct reliable XC potentials by properly mixing the exact exchange and the local density approximation potentials in real space. The spatially dependent mixing parameter is derived based on the derivative discontinuity of electron density and is first-principle. We derived the equations for solving the mixing parameter and proposed an approximation to simplify these equations. Based on this approximation, this new method gives reasonable predictions for the ionization energies, fundamental gaps, and singlet-triplet energy differences for various molecular systems. The impact of the approximation on the constructed XC potentials is examined, and it is found that the quality of the XC potentials can be further improved by removing the approximation. This work demonstrates that the derivative discontinuity of electron density is a promising constraint for constructing high-quality XC potentials.

    \end{abstract}

  %\pacs{Valid PACS appear here}% PACS, the Physics and Astronomy
  % Classification Scheme.
  %\keywords{density functional theory, exact exchange, exchange-correlation, optimized effective potential}
  %Use showkeys class option if keyword
    %display desired

\maketitle
\thispagestyle{firstpage}

\section{Introduction}

In Kohn-Sham density functional theory (KS-DFT),\cite{Hohenberg1964,Kohn1965} the accuracy of electronic structure predictions is determined by the quality of the exchange-correlation (XC) potential. Much effort has been devoted to improving the XC energy functionals.\cite{Perdew2001,Scuseria2005} 
Among them, hybrid functionals\cite{Becke1993a} have gained much interest due to their ability to reduce the self-interaction error\cite{Perdew1981,Zhang1998,MoriSanchez2006,Cohen2011} by incorporating the exact exchange (EXX) in the functionals.
The accuracy of the hybrid functionals depends on the mixing parameter, which should be system-dependent. For example, the connection between the mixing parameter and the dielectric function has been recognized.\cite{Shimazaki2008,Shimazaki2009,Shimazaki2010,Alkauskas2011,Marques2011,Moussa2012,RefaelyAbramson2013,Koller2013,Skone2014} 
%------------------------
More generally, the mixing parameter should be spatially dependent. For instance, it should be one in one-electron regions. Spatially dependent mixing parameters have been actively developed in the field of local hybrid functionals\cite{Cruz1998,Jaramillo2003,Haunschild2009,Maier2018}. For example, the mixing parameter has been formulated based on the ratio between the von Weizs{\"a}cker kinetic energy density and the KS kinetic energy density,\cite{Jaramillo2003,Arbuznikov2006,Bahmann2007,Perdew2008,Schmidt2014} the reduced density gradient,\cite{Arbuznikov2007,Kaupp2007,Arbuznikov2009,Schmidt2014} an indicator that measures the exponential decay of electron densities,\cite{Silva2015} and the local dielectric function.\cite{Borlido2018,Zheng2019,Zhan2023}

Above mentioned works mainly focused on improving the XC energy functional. In this work, we focus on improving the XC potential in hybrid DFT. We develop a method to fully determine the spatially dependent parameter for mixing the EXX and local density approximation (LDA) potentials. 
The method is termed spatial mixing of model potentials (SMMP). 
SMMP is inspired by the optimally tuned range-separated hybrid functionals (OT-RSH),\cite{Stein2009,Baer2010,Stein2010,Kronik2012} in which the range-separation parameter is optimized based on the ionization potential (IP) theorem.\cite{Perdew1982,Perdew1983,Levy1984,Perdew1997} A limitation of the IP theorem is that it only provides one equation, making it impossible to determine the mixing parameter in real space. 
Our method employs a new condition: electron density changes linearly between states with $N$ and $N-1$ electrons.\cite{Perdew1982} 
The number of equations given by this condition is equal to the number of grid points, making it possible to fully determine the mixing parameter in real space. 
%-----------------------
Compared to OT-RSH, SMMP faces several challenges. First, the EXX potentials used for the potential mixing are calculated using the optimized effective potential (OEP) method,\cite{Sharp1953,Talman1976,Kuemmel2008} which is numerically more complicated than the generalized KS method used by OT-RSH. 
Secondly, while solving the mixing parameter, we need to invert a system's KS linear response function, which is another OEP problem. 
Thirdly, SMMP is currently not applicable to systems with degenerate highest occupied molecular orbitals (HOMOs). These challenges will be discussed in this work. Despite these challenges, this work demonstrates a new, promising direction to construct reliable XC potentials for hybrid DFT.

We note that the idea of improving the XC potential is not new. In the field of the time-dependent DFT,\cite{Runge1984,Marques2004} much effort has been devoted to improving the XC potential to achieve better predictions of excitation energies.\cite{Casida1998,Casida2000,Tozer1998,Handy1999,Gritsenko1999,Schipper2000} 
In these methods, the correct asymptotic behavior of the XC potential at $r \rightarrow \infty$ is ensured by mixing local/semilocal XC potentials with the model potentials that have the correct asymptotic behavior. The mixing parameters were usually calculated based on certain models. In this work, we focus on building reliable XC potentials for ground-state electronic structure calculations, and the mixing parameter is fully first-principle.

The paper is organized as follows. First, we derive the equations for calculating the parameter for mixing the EXX and LDA potentials. We then introduce an approximation to simplify these equations. The performance of SMMP is demonstrated by calculating the ionization energies, fundamental gaps, and singlet-triplet energy differences of various molecular systems. In the end, the impact of the orbital relaxation on the mixing parameter is examined.

\section{Theoretical Methods} \label{sec:theory}

In KS-DFT, the KS potential is defined as
\begin{equation} \label{eq:vs}
	v_{s,\sigma}(\vec r) = v_H(\vec r) + v_{xc,\sigma}(\vec r) + v_{ext}(\vec r)
\end{equation}
where $\sigma$ is the spin index, $v_H$ is the Hartree potential, and $v_{ext}$ is the external potential. In this work, the XC potential is constructed by mixing the EXX potential $v^{EXX}_{x,\sigma}$ and the LDA XC potential $v^{LDA}_{xc,\sigma}$ as  
\begin{equation} \label{eq:vxc}
	v_{xc,\sigma}(\vec r) = f(\vec r) v^{EXX}_{x,\sigma}(\vec r) + (1-f(\vec r)) v^{LDA}_{xc,\sigma}(\vec r),
\end{equation}
where $f(\vec r)$ is the mixing parameter. For the LDA functional, the Perdew-Wang parametrization is employed.\cite{Perdew1992,Perdew2018} 
For simplicity, we only consider the case that the mixing parameter is the same for different spins. An extension to spin-polarized mixing parameters will be given in a future work. Different from conventional hybrid DFT calculations in which the generalized KS theory is employed, $v_{x,\sigma}^{EXX}$ in Eq.~\ref{eq:vxc}  is obtained by solving the OEP equation. This makes it straightforward to mix the LDA and EXX potentials. In this work, we solve the OEP equation using the Krieger-Li-Iafrate (KLI) approximation.\cite{Krieger1992}

We have also examined the performance of SMMP by replacing the LDA potential in Eq.~\ref{eq:vxc} with the Perdew-Burke-Ernzerhof (PBE) XC potential\cite{Perdew1996}. The HOMO eigenvalues obtained from the PBE mixing are very close to those obtained from the LDA mixing. This suggests that the mixing parameter can automatically adjust itself for different density functionals to satisfy the derivative discontinuity condition of the electron density.

For a system with $N-a$ ($0<a<1$) electrons, its electron density is a linear combination of two systems with $N$ and $N-1$ electrons
\begin{equation}
	\rho_{N-a,\sigma}(\vec r) = (1-a)\rho_{N,\sigma}(\vec r) + a \rho_{N-1,\sigma}(\vec r).
\end{equation} 
The above equation leads to the following condition
\begin{equation}\label{eq:L1}
	\frac{d \rho_{N,\sigma}(\vec r) }{d N} = \rho_{N,\sigma}(\vec r) 
	- \rho_{N-1,\sigma}(\vec r ),
\end{equation}
in which the derivative is taken from the left side of $N$. All terms in Eq.~\ref{eq:L1} are functionals of $f(\vec r)$, and in principle we can determine $f(\vec r)$  by minimizing the following
\begin{equation}
	G = \int \sum_{\sigma=\alpha, \beta} 
	\left(\frac{d \rho_{N,\sigma}(\vec r) }{d N}  - (\rho_{N,\sigma}(\vec r) 
	- \rho_{N-1,\sigma}(\vec r))  \right)^2 d r^3 .
\end{equation}
However, this minimization is difficult in practice, since $G$ is an implicit functional of $f(\vec r)$, making it difficult to calculate the gradient $\delta G/\delta f(\vec r)$. To overcome this difficulty, in what follows we derive an iterative scheme to solve the mixing parameter.

Let's consider a spin density $\rho_{\sigma}$, we then have
\begin{equation} \label{eq:1}
	\frac{d \rho_\sigma(\vec r)}{d N} = 
	\left. \frac{\partial \rho_\sigma(\vec r) }{\partial N}\right|_{v_{s,\sigma}} + 
	\int \chi_{s,\sigma}(\vec r, \vec r')
	\frac{d v_{s,\sigma} (\vec r')}{d N} dr'^3
\end{equation}
where the derivative is taken on the left-hand side of $N$. $\chi_{s,\sigma} = \delta \rho_{\sigma} / \delta v_{s,\sigma}$ is the KS linear response function for spin $\sigma$. 
The linearity condition in Eq.~\ref{eq:L1} can then be imposed by  replacing $d\rho_\sigma/dN$ in Eq.~\ref{eq:1} with  $\rho_{N,\sigma}-\rho_{N-1,\sigma}$. Also, to simplify the notations, let's define 
\begin{equation}\label{eq:gamma}
	\gamma_\sigma = 
	\rho_{N,\sigma}-\rho_{N-1,\sigma} - 
	\left. \frac{\partial \rho_\sigma}{\partial N}\right|_{v_{s,\sigma}}. 
\end{equation}
$\gamma_\sigma$ measures the orbital relaxation after removing one electron. 
If the HOMO is from spin $\alpha$, we have $\gamma_\alpha=\rho_{N,\alpha}-\rho_{N-1,\alpha}-(\phi_{\alpha}^\mathrm{HOMO})^2$ and  $\gamma_\beta=\rho_{N,\beta}-\rho_{N-1,\beta}$, where  $\phi_{\alpha}^\mathrm{HOMO}$ is the HOMO. For cases where the HOMOs of spin $\alpha$ and spin $\beta$ have the same eigenvalue, we can choose HOMO from either spin $\alpha$ or spin $\beta$.

From Eq.~\ref{eq:1}, $dv_{s,\sigma}/dN$ can be obtained by inverting $\chi_{s,\sigma}$ 
\begin{equation}   \label{eq:dv}
	\frac{d v_{s,\sigma} (\vec r)}{d N} = 	\int  \chi_{s,\sigma}^{-1}(\vec r,\vec r') \gamma_\sigma(\vec r') 
	d r'^3. 
\end{equation}
In general, $dv_{s,\sigma}/dN$ can only be determined up to an unknown constant due to the inversion of $\chi_{s,\sigma}$. However, this constant can be fixed by requiring $dv_{s,\sigma}/dN \rightarrow 0$ in the vacuum, which is required for deriving the correct asymptotic behavior for the mixing parameter in Eq.~\ref{eq:fasym}. Therefore, Eq.~\ref{eq:dv} is the correct expression for  $dv_{s,\sigma}/dN$.

On the other hand, by taking the derivative of $v_{s,\sigma}$ with respect to $N$ in Eq.~\ref{eq:vs}, we obtain
\begin{equation}\label{eq:dvdN3}
	\frac{dv_{s,\sigma}(\vec r)}{d N} 
	= h_\sigma(\vec r) + f(\vec r) g_\sigma(\vec r)
\end{equation}
with $h_\sigma$ and $g_\sigma$ defined as
\begin{eqnarray}
	\label{eq:h}
	&& h_\sigma(\vec r) = \frac{dv_H(\vec r)}{dN} + \frac{dv_{xc,\sigma}^{LDA}(\vec r)}{dN} \\
	\label{eq:g} 
	&& g_\sigma(\vec r) = \frac{dv_{x,\sigma}^{EXX}(\vec r)}{dN} -  \frac{dv_{xc,\sigma}^{LDA}(\vec r)}{dN}. 
\end{eqnarray}
Combining Eq.~\ref{eq:dv} and Eq.~\ref{eq:dvdN3}, we obtain the key equation for the mixing parameter
\begin{equation} \label{eq:alpha1}
	h_\sigma(\vec r) + f(\vec r) g_\sigma(\vec r) = p_\sigma(\vec r)
\end{equation}
where  $p_\sigma$ is the shorthand for 
\begin{equation} \label{eq:p}
	p_\sigma(\vec r) = 
	\int \chi_{s,\sigma}^{-1}(\vec r,\vec r') \gamma_\sigma(\vec r') d r'^3. 
\end{equation}
Since generally we  cannot find a single $f(\vec r)$ to satisfy Eq.~\ref{eq:alpha1} for both spins,  the optimal $f(\vec r)$ is obtained by minimizing the cost function
\begin{equation}
	W = \sum_\sigma 
	\int [h_\sigma(\vec r) + f(\vec r) g_\sigma(\vec r) - p_\sigma(\vec r)]^2  d r^3 .
\end{equation}
By setting $\delta W / \delta f=0$, we reach the final expression for the mixing parameter
\begin{equation}\label{eq:f}
	f(\vec r) = \frac{\sum_\sigma (p_\sigma(\vec r) 
		- h_\sigma(\vec r))g_\sigma(\vec r)}{\sum_\sigma g^2_\sigma(\vec r)}.
\end{equation}

For a molecular system, the XC potential should reduce to the EXX potential in the vacuum, that is, $v_{xc} \sim -1/r$ for $r\rightarrow \infty$.\cite{Gunnarsson1979,Levy1984,Almbladh1985} This suggests the following asymptotic behavior for the mixing parameter
\begin{equation}\label{eq:fasym}
	f(\vec r) \rightarrow 1 \mathrm{\ for\ } r \rightarrow \infty.
\end{equation}
In what follows, we show that the above condition is ensured by Eq.~\ref{eq:f}. 
Let's assume that the HOMO is from spin $\alpha$.  A change of the electron number will only cause the spin-$\beta$ orbitals to relax, and therefore $d v_{x,\beta}^{EXX} / dN$ decays faster than $1/r$ at $r \rightarrow \infty$. $d v_{xc,\beta}^{LDA}/dN$ also decays exponentially due to the exponential decay of the electron density. Based on these, $g_\beta$ defined in Eq.~\ref{eq:g} decays faster than $1/r$ in the vacuum. On the other hand, $p_\sigma$ (Eq.~\ref{eq:p}) decays faster than $1/r$ because the integration of $\gamma_\sigma$ is zero. For the spin-$\alpha$ channel, we have $h_\alpha \approx dv_H/dN$ and $g_\alpha \approx dv_{x,\alpha}^{EXX}/dN$ in the vacuum, and therefore $h_\alpha$ and $g_\alpha$ both decay as $1/r$.  Eq.~\ref{eq:f} then reduces to $-h_\alpha/g_\alpha$ in the vacuum. Also note that $dv_{x,\alpha}^{EXX}/dN \approx -dv_H/dN$ in the vacuum. The asymptotic condition in Eq.~\ref{eq:fasym} is then proved.

One major numerical challenge is the calculation of $p_\sigma$ defined in Eq.~\ref{eq:p}, due to the inversion of $\chi_{s,\sigma}$. 
This is a well-known problem in the field of OEP.\cite{Hirata2001,HeatonBurgess2007} 
To avoid calculating $p_\sigma$ in this work, we introduce a frozen orbital approximation (FOA). A full calculation of $p_\sigma$ will be explored in future work. Under FOA, all orbitals are assumed to not relax after removing one electron from the system. This gives $\frac{d \rho_\sigma(\vec r)}{d N} =
\left. \frac{\partial \rho_\sigma(\vec r) }{\partial N}\right|_{v_{s,\sigma}}$ which leads to $p_\sigma=0$. Eq.~\ref{eq:f} then reduces to 
\begin{equation}\label{eq:fFOA}
	f^{FOA}(\vec r) = - \frac{\sum_\sigma h_\sigma(\vec r) g_\sigma(\vec r)}{\sum_\sigma g^2_\sigma(\vec r)}.
\end{equation}
By a similar argument, we can show that $f^{FOA}$ has the same asymptotic behavior as in Eq.~\ref{eq:fasym}.

Next, we discuss the calculations of $h_\sigma$ and $g_\sigma$ defined in Eqs.~\ref{eq:h} and \ref{eq:g}. They rely on $dv_H/dN$, $dv_{x,\sigma}^{EXX}/dN$, and $dv_{xc,\sigma}^{LDA}/dN$. In this work, these derivatives are calculated using the second-order finite-difference method by evaluating $v_H$, $v_{x,\sigma}^{EXX}$, and $v_{xc,\sigma}^{LDA}$ at $N$, $N-\delta$, and $N-2\delta$ electron numbers, with $\delta=0.05$. In general, we should perform self-consistent KS-DFT calculations at these electron numbers. But under FOA we can simply take the orbitals from the $N$-electron calculations and only modify the occupation numbers for the $N-\delta$ and $N-2\delta$ systems. This greatly simplifies the calculations. All results in this work are calculated using FOA, if not specified.

One limitation of SMMP is that it is currently only applicable to systems without degenerate HOMOs. This is because the method removes a small amount of electron from the HOMO for calculating $h_\sigma$ and $g_\sigma$. For systems with degenerate HOMOs, the selection of one HOMO from multiple degenerate HOMOs may result in a mixing parameter with a symmetry different from that of the system. We are exploring several ways to extend SMMP to systems with degenerate HOMOs.

The iterative scheme for solving the mixing parameter is as follows. Starting with a trial mixing parameter, a self-consistent KS-DFT calculation is performed with the XC potential mixed based on Eq.~\ref{eq:vxc}. $h_\sigma$, $g_\sigma$, and $p_\sigma$ are then calculated based on Eqs.~\ref{eq:h}, \ref{eq:g}, and \ref{eq:p}, respectively. Under FOA, $p_\sigma$ is set to zero. The mixing parameter is then updated using Eq.~\ref{eq:f}. Under FOA, the mixing parameter is updated using Eq.~\ref{eq:fFOA}. We then check the convergence of the mixing parameter. If it is converged, the program exits. Otherwise, we go to the first step. In this work, the convergence of the mixing parameter is accelerated using the Pulay mixing\cite{Pulay1980}. The convergence is usually fast.

\section{Numerical Details}

This new method is implemented in both ABINIT\cite{Gonze2020,Romero2020} and PySCF programs\cite{Sun2015,Sun2017,Sun2020}. The structures of the molecular systems are optimized using the ORCA program\cite{Neese2011,Neese2017,Neese2020}.  H$_2$ and CO in Section~\ref{sec:mix} are calculated using the modified PySCF program with the cc-pVTZ basis set.\cite{Dunning1989,Kendall1992,Woon1993} H$_2$'s bond length is 0.74 \AA, and CO's bond length is 1.128 \AA. The tests in Section~\ref{sec:IE} are performed using the modified ABINIT program. Molecular structures are plotted using the VESTA program.\cite{Momma2011}

For the ionization energy tests, the following molecules are studied: Na$_4$, butane, phenol, chloro-benzene, thiophene, acetone, glyoxal, formic acid, formyl fluoride, H$_2$O, N$_2$, H$_2$, NH$_3$, CO, propane, methylamine, methylene imine, pyrrole, ethanol, ethene, formaldehyde, and $\mathrm{K_2O}$.  Their structures are relaxed using the B3LYP functional\cite{Becke1993,Lee1988,Vosko1980,Stephens1994} and 6-311G** basis sets\cite{McLean1980,Petersson1991}.
The radicals are prepared based on these molecules, and their molecular formulas are CH$_3$CHCH$_2$CH$_3$, CH$_3$CH$_2$O, FCO, C$_2$H$_3$, CH$_2$N, NH$_2$, OH, HCO, CH$_3$CHCH$_3$, CH$_3$NH, COOH, HCOO, C$_6$H$_5$O, C$_4$H$_3$S, C$_4$H$_4$N, CHOCO, and C$_6$H$_4$Cl (the H attached to the third carbon is removed). All radicals are relaxed using M{\o}ller–Plesset perturbation theory\cite{Moeller1934} to the second order (MP2) with the cc-pVTZ basis. 
Anions are prepared by adding one electron to these radicals, except OH which is not considered due to the degenerate HOMOs. We also include three other anions: CH$_2$CH$^-$, CH$_2^-$, and $\mathrm{CH_3^-}$. All anions are relaxed using MP2 with the aug-cc-pVTZ basis.
The benchmarks for the ionization energies are obtained using the coupled cluster method with single, double, and perturbative triple excitations\cite{Cizek1966,Cizek1969,Paldus1972,Paldus1977,Yarkony1995}  (CCSD(T)) for $N$ and $N-1$ electrons using the ORCA program. Basis extrapolation is used to estimate CCSD(T) energies at the complete basis set limit. For some large systems, extrapolation cannot be employed due to the memory limit. Details about the CCSD(T) calculations are given in the Supplementary Material.
The ABINIT-based calculations are performed using a kinetic energy cutoff of 1000 eV and the Goedecker-Teter-Hutter (GTH) pseudopotentials.\cite{Goedecker1996,Hartwigsen1998,Krack2005} 
With GTH pseudopotentials, ABINIT can calculate the Hartree potentials in real space using a Poisson solver.\cite{Genovese2007} This is necessary for the finite-difference calculations of $dv_H/dN$, for which a small amount of electron is removed from the system. The integrals in the EXX energy and potential calculations are also calculated in real space using the Poisson solver.

For the fundamental gap tests, we employ the following molecules: 1-chloro-2-nitrobenzene, 1-nitronaphthalene, benzoquinone, 1,4-naphthoquinone, azulene, Na$_4$, and glyoxal. Their structures are relaxed using the cc-pVDZ basis sets and the B3LPY functional.
The benchmarks are from  CCSD(T)\cite{Cizek1966,Cizek1969,Paldus1972,Paldus1977,Yarkony1995} calculations. Since anions are involved, diffuse basis functions are used for all calculations. The CCSD(T) energies of Na$_4$ and glyoxal are extrapolated based on the aug-cc-pVDZ and aug-cc-pVTZ basis sets. Other molecules are calculated using the aug-cc-pVDZ basis due the memory limitation. For the singlet-triplet energy gap calculations, CCSD(T) energies are obtained through extrapolation with the details given in the Supplementary Material, except naphthalene which is calculated using the cc-pVDZ basis.

\section{Results and Discussions}

\subsection{Tests on ionization energies, fundamental gaps, and singlet-triplet energy differences}\label{sec:IE}

\begin{figure}[]
	\centering
	\includegraphics[width=0.5\textwidth]{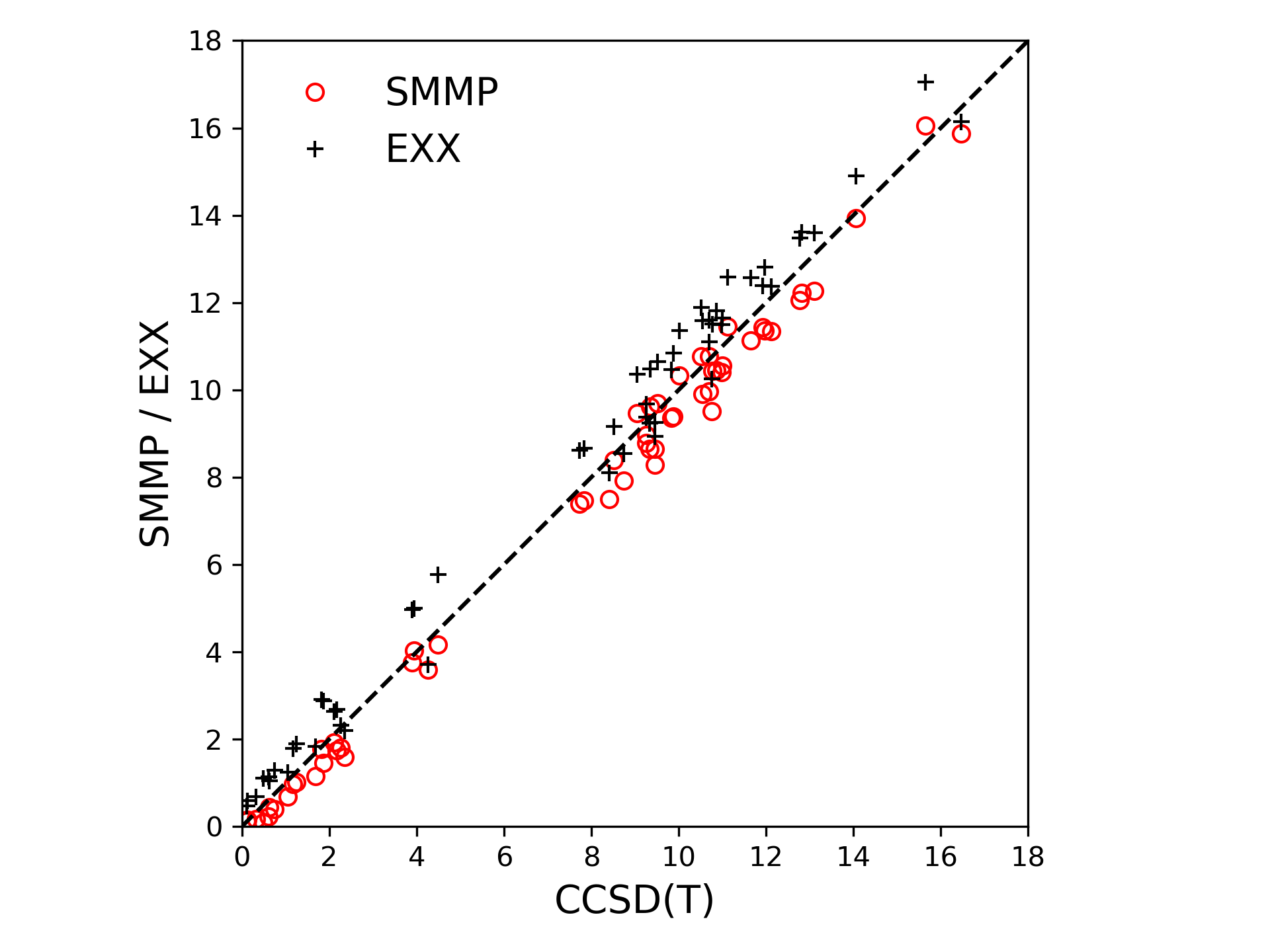}
	\caption{Compare the ionization energies (in eV) from CCSD(T), SMMP (based on FOA), and KS-DFT-EXX calculations. }
	\label{fig:IE}
\end{figure}

We first examine the performance of SMMP (based on FOA) by calculating the vertical ionization energies for various molecules, radicals, and anions. The negative HOMO eigenvalues are taken as the ionization energies. The results are summarized in Fig.~\ref{fig:IE}. CCSD(T) results are considered as the benchmarks and are obtained by removing one electron and then calculating the energy change. For comparison, we also show the negative HOMO eigenvalues from KS-DFT EXX calculations. As expected, EXX generally produces more negative HOMO eigenvalues, with most data points above the dashed line. SMMP can address this problem, but it tends to overcorrect it. Most of the SMMP results are below the dashed line. Such overcorrection can also be observed in Table~\ref{tb:IE_mae} and Figure 1 in the Supplementary Materials. Table~\ref{tb:IE_mae} shows that the mean errors from EXX calculations are all positive, whereas the mean errors from SMMP calculations are all negative.
SMMP corrects EXX's HOMO eigenvalues by generating positive KS correlation potentials. In Fig.~\ref{fig:H2}(c), SMMP predicts a positive KS correlation potential for H$_2$ over the entire system. For CO in Fig.~\ref{fig:CO}(c), SMMP predicts a KS correlation potential that is positive in the region where the orbitals decay.

\begin{table}
	\caption{Mean absolute errors (MAE) and mean errors (ME) (in eV) of the ionization energies of the molecular systems in Fig.~\ref{fig:IE} calculated by SMMP (based on FOA) and KS-DFT-EXX.}
	\centering
	\begin{tabular}{lcccc}
		\hline\hline
		& \multicolumn{2}{c}{SMMP} & \multicolumn{2}{c}{EXX} \\ \cline{2-3}\cline{4-5}
		& MAE  &        ME         & MAE  &        ME        \\ \hline
		molecule & 0.57 &       -0.53       & 0.68 &       0.44       \\
		radical  & 0.45 &       -0.25       & 0.82 &       0.80       \\
		anion    & 0.28 &       -0.27       & 0.55 &       0.53       \\ \hline\hline
	\end{tabular}	\label{tb:IE_mae}
\end{table}

The accuracy of SMMP and EXX can be examined based on the mean absolute errors (MAE), as given in Table~\ref{tb:IE_mae}. We calculate MAEs for molecules, radicals, and anions, separately. For molecules, SMMP's MAE is only slightly smaller than that of EXX. The errors in these SMMP calculations could be due to the use of FOA. When an electron is removed from these molecules, the orbital relaxations can be significant sometimes, which then leads to large errors in these FOA-based SMMP calculations. 
It is interesting to see that, for radicals and anions, SMMP's MAEs are considerably smaller than those of EXX.  The reason could be that, in the case of a radical, the removal of an unpaired electron causes less orbital relaxation compared to the case of molecules. As a result, FOA-based SMMP performs better for radicals. Similarly, for an anion, the removal of the extra electron has a smaller impact on the orbitals compared to the case of molecules, which makes FOA-based SMMP perform better.

\begin{figure}[]
	\centering
	\includegraphics[width=0.5\textwidth]{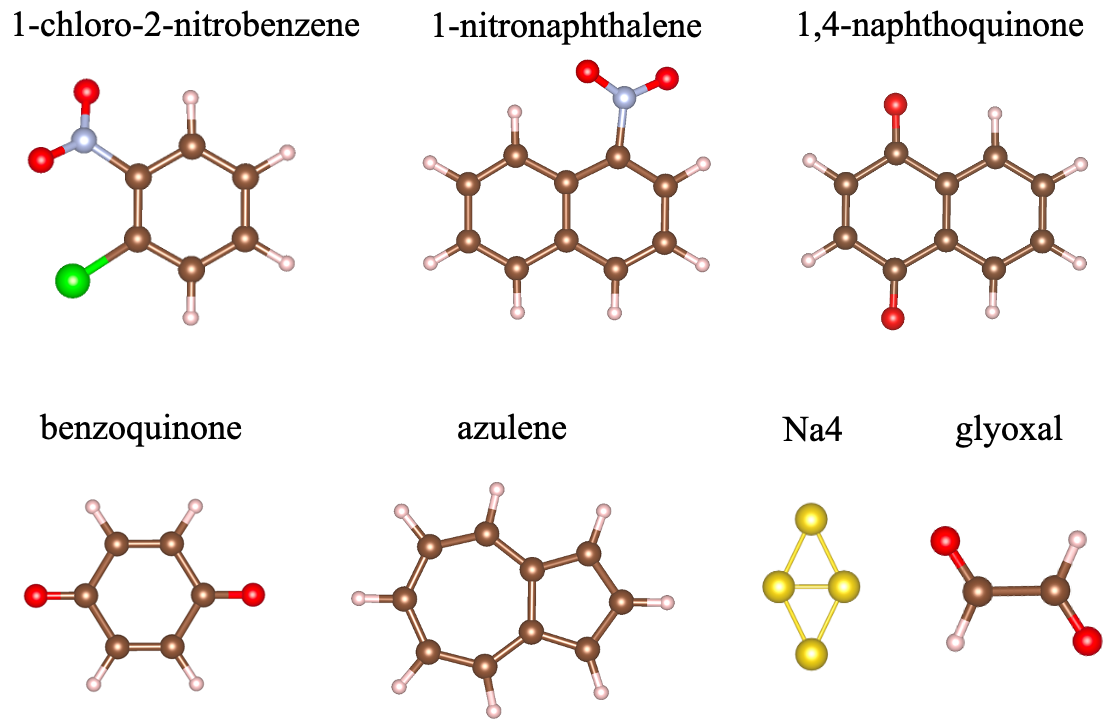}
	\caption{Molecules for fundamental gap calculations.}
	\label{fig:mol}
\end{figure}

\begin{figure}[]
	\centering
	\includegraphics[width=0.5\textwidth]{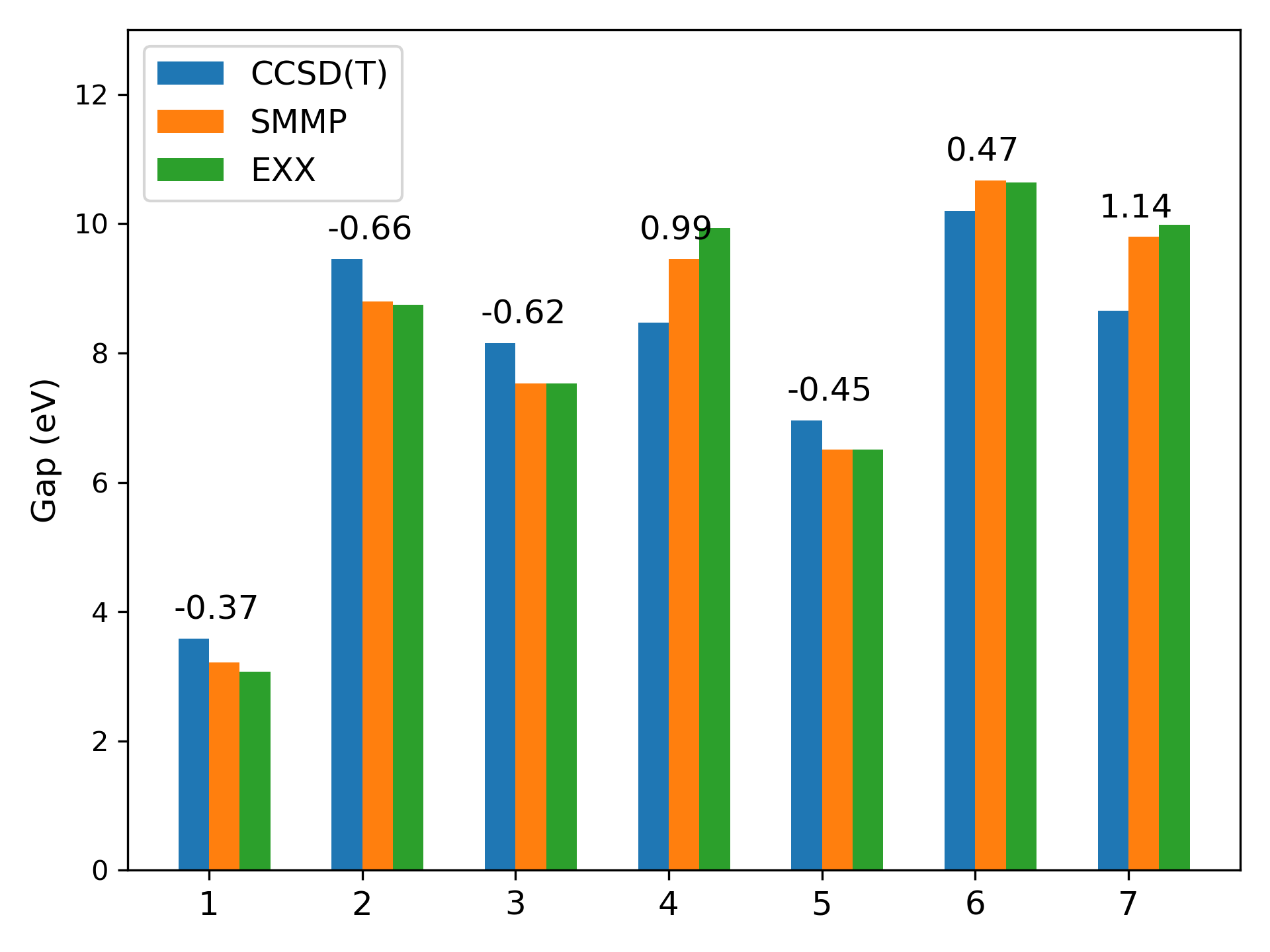}
	\caption{Fundamental gaps from CCSD(T), SMMP, and KS-DFT EXX calculations. 1: Na$_4$, 2: 1-chloro-2-nitrobenzene, 3: 1-nitronaphthalene, 4: 1,4-naphthoquinone, 5: azulene, 6: glyoxal, and 7: benzoquinone. The errors (SMMP - CCSD(T), in eV) are at the top of each bar.}
	\label{fig:gap}
\end{figure}

\begin{figure}[]
	\centering
	\includegraphics[width=0.5\textwidth]{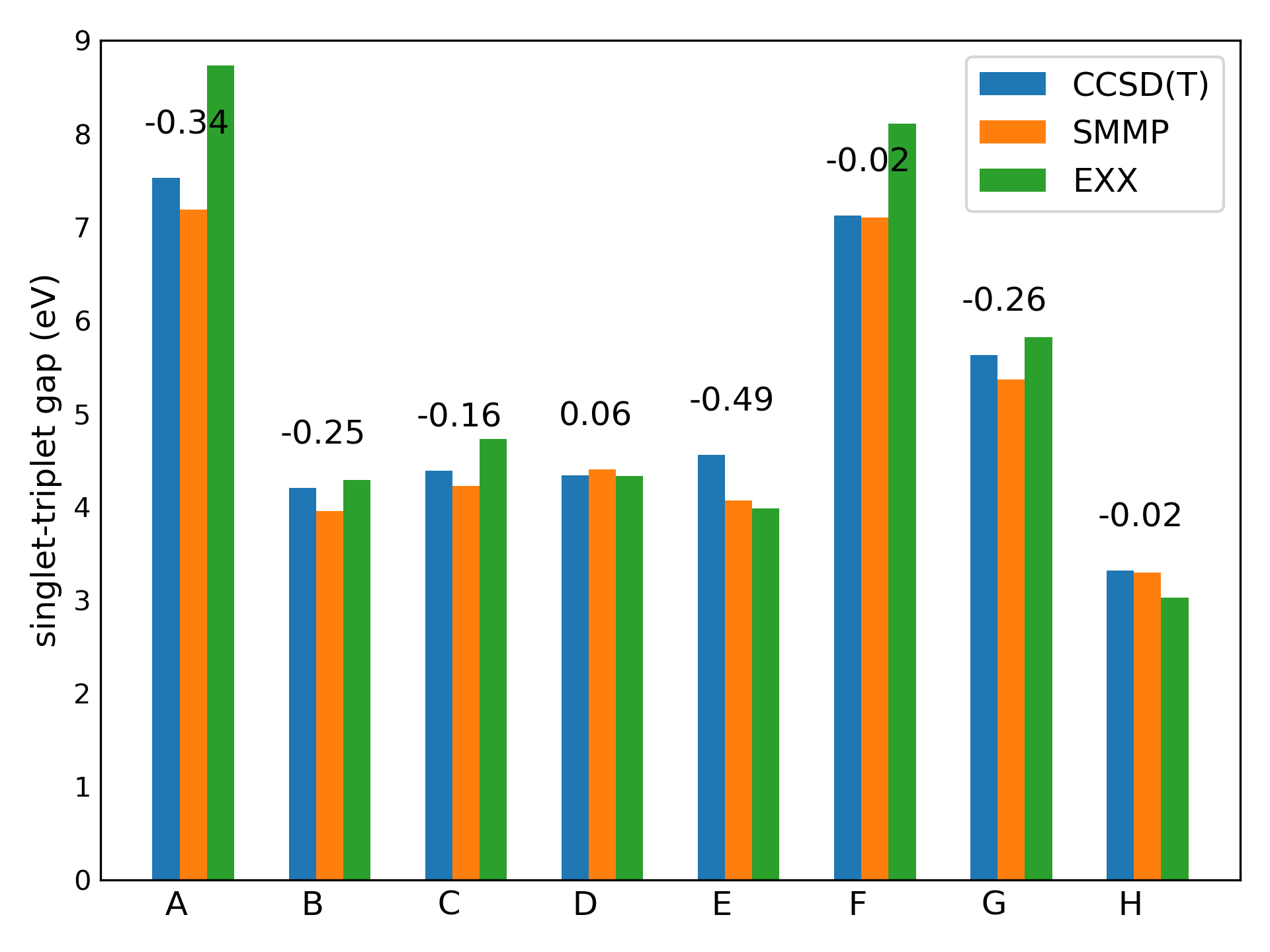}
	\caption{The singlet-triplet energy differences from CCSD(T), SMMP, and KS-DFT EXX calculations.  A: water. B: aceton. C: Cl-benzene. D: phenol. E: pyrrole. F: ethanol. G: formic acid. H: naphthalene. The errors (SMMP - CCSD(T), in eV) are at the top of each bar.}
	\label{fig:st_gap}
\end{figure}

Next, we examine SMMP's performance for predicting the fundamental gaps and the singlet-triplet (ST) energy differences of several molecules. The molecules for the fundamental gap calculations are shown in Fig.~\ref{fig:mol}.
The results are summarized in Fig.~\ref{fig:gap} and Fig.~\ref{fig:st_gap}. The benchmarks are from CCSD(T) calculations. For CCSD(T), the fundamental gaps are calculated as 
$E_\mathrm{CCSD(T)}^{N-1} + E_\mathrm{CCSD(T)}^{N+1} - 2 E_\mathrm{CCSD(T)}^{N}$. For SMMP, the fundamental gap is calculated as $\epsilon_{N+1}^\mathrm{HOMO} - \epsilon_{N}^\mathrm{HOMO}$, with $\epsilon_{M}^\mathrm{HOMO}$ being the eigenvalue of the HOMO for an $M$-electron system. For the ST energy difference, the triplet's geometry is the same as the singlet state. For SMMP, the ST energy difference is calculated as $\epsilon_\mathrm{triplet}^\mathrm{HOMO}-\epsilon_\mathrm{singlet}^\mathrm{HOMO}$.

Fig.~\ref{fig:gap} and Fig.~\ref{fig:st_gap} show that the SMMP results agree reasonably with the benchmarks for most systems. We also note that EXX performs similarly as SMMP for many systems. This is somehow surprising, since EXX typically predicts too negative eigenvalues for HOMOs. The seemly good performance of EXX is actually due to error cancellations. 
For the fundamental gaps, EXX gives too negative HOMO eigenvalues for both $N$ and $N+1$ electron systems, which leads to an effective error cancellation for the calculations of  $\epsilon_{N+1}^\mathrm{HOMO} - \epsilon_{N}^\mathrm{HOMO}$. For the ST gaps, EXX also predicts too low eigenvalues for the HOMOs of the singlet and triplet states, which leads to similar error cancellations for  calculating $\epsilon_\mathrm{triplet}^\mathrm{HOMO}-\epsilon_\mathrm{singlet}^\mathrm{HOMO}$. Note that these error cancellations do not always work. For example, EXX gives much error in the ST energy differences for water and ethanol (Fig.~\ref{fig:st_gap}). Overall, SMMP outperforms EXX in all these tests.

\subsection{Mixing parameters based on FOA and the impact of orbital relaxation}\label{sec:mix}

In what follows, we examine the mixing parameters calculated based on FOA. As our long-term goal is to construct accurate XC potentials using SMMP, we especially examine on whether incorporating orbital relaxation into the mixing parameter calculations can improve the accuracy of XC potentials. This is achieved by analyzing how orbital relaxation affects the electron densities predicted by SMMP.

\subsubsection{H$_2$}

\begin{figure}[]
	\includegraphics[width=0.5\textwidth]{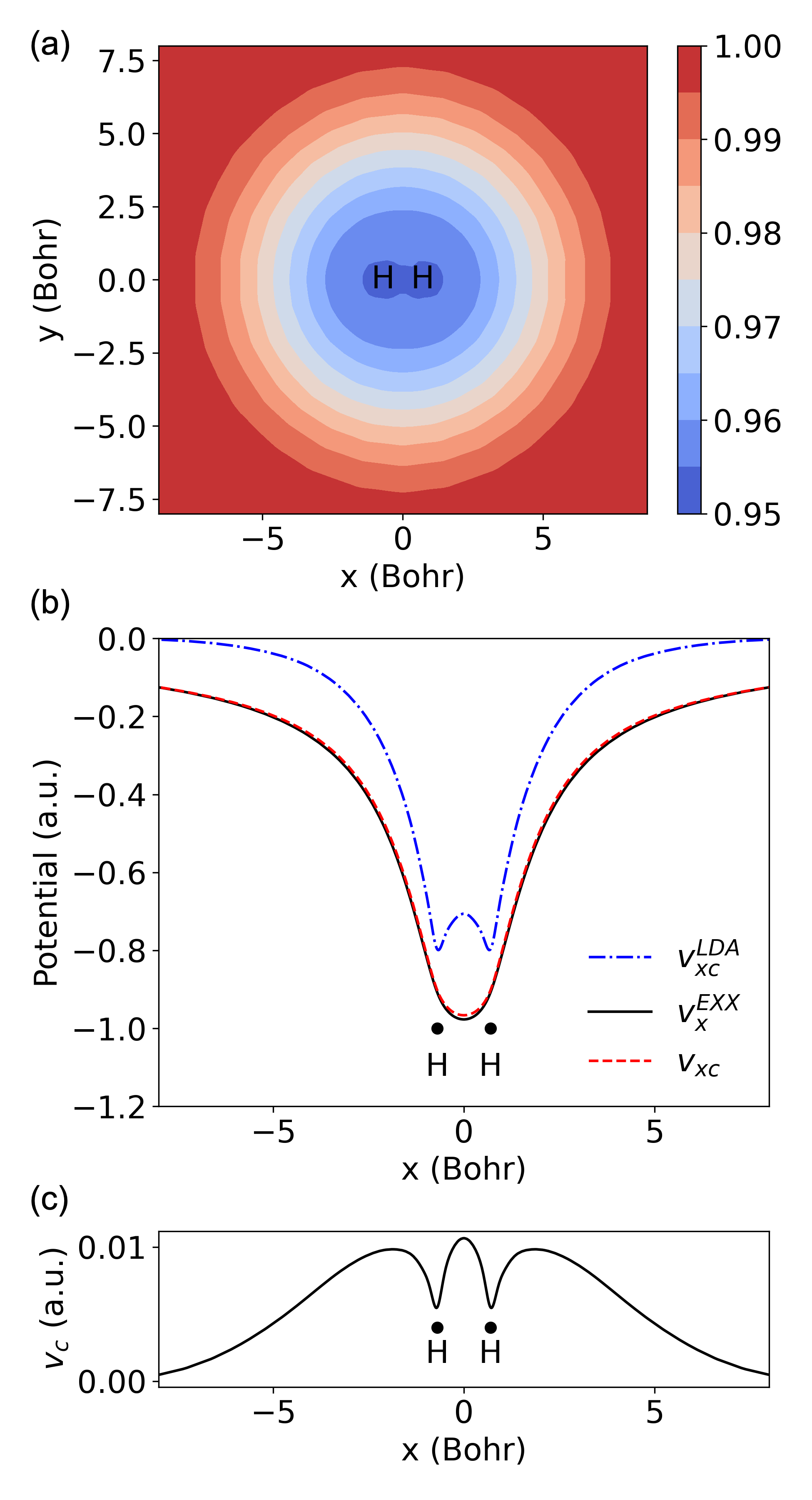}
	\caption{(a) Contour plot of the mixing parameter of H$_2$ on the plane of the two H atoms, calculated based on FOA. (b) Comparison of LDA, EXX, and SMMP XC potentials of H$_2$.}
	\label{fig:H2}
\end{figure}

The mixing parameter calculated based on FOA is given in Fig.~\ref{fig:H2}(a). As expected, it approaches one in the vacuum, confirming the condition in Eq.~\ref{eq:fasym}.
As a result, $v_{xc}$ correctly reduces to $v_x^{EXX}$ in vacuum (Fig.~\ref{fig:H2}(b)). This is consistent with the exact XC potential obtained in a recent work.\cite{Kanungo2019} In Fig.~\ref{fig:H2}(b), another interesting observation is that $v_{xc} > v_{x}^{EXX}$ near the vacuum, which gives a positive KS correlation potential $v_c$ near the vacuum (see Fig.~\ref{fig:H2}(c)). This is also consistent with the previous work.\cite{Kanungo2019}

One problem is that the mixing parameter is smaller than one around the atoms, which leads to $v_{xc} > v_x^{EXX}$. This in turn produces a positive $v_c$ around the atoms, which disagrees with the exact results from the previous works.\cite{Gritsenko1995,Kanungo2019} The exact $v_c$ should be negative around the hydrogen atoms and is only positive near the middle of the bond. In what follows, we show that this problem should be due to the use of FOA in these SMMP calculations.

To examine the impact of FOA on the mixing parameter, we perform a different calculation in which $h_\sigma$ and $g_\sigma$ are calculated using relaxed KS orbitals. This means that  $dv_H/dN$, $dv_{x}^{EXX}/dN$, and $dv_{xc}^{LDA}/dN$ in Eq.~\ref{eq:h} and Eq.~\ref{eq:g} are calculated by performing self-consistent KS-DFT calculations at $N-\delta$ and $N-2\delta$ electrons. $p_\sigma$ is still set to zero. 
With these settings, we find that H$_2$'s HOMO eigenvalue is predicted to be too high. This indicates that the mixing scheme in Eq.~\ref{eq:vxc} should be revised when the orbital relaxation is considered. 
After trying several different mixing schemes, we find that good predictions for the HOMO eigenvalues can be obtained using the following new mixing scheme
%------------------------------------
\begin{equation}\label{eq:new_mix}
	v_{xc,\sigma}(\vec r) = 
	f'(\vec r) v_{x,\sigma}^{EXX}(\vec r) + 
	(1-f'(\vec r)) v_{x,\sigma}^{LDA}(\vec r) + 
	v_{c,\sigma}^{LDA}(\vec r).
\end{equation}
%------------------------------------
This new mixing scheme is actually consistent with conventional hybrid functionals, in which the local/semilocal correlation energies are often not mixed. The mixing parameter for this new mixing scheme can be derived in a similar way as discussed in Section~\ref{sec:theory}
%------------------------
\begin{equation}\label{eq:f_newmix}
	f'(\vec r) = \frac{\sum_\sigma (p_\sigma(\vec r) - h'_\sigma(\vec r))g'_\sigma(\vec r)}{\sum_\sigma g'^2_\sigma(\vec r)},
\end{equation}
%------------------------
with $h'_\sigma$ and $g'_\sigma$ defined as 
\begin{eqnarray}
	\label{eq:h_new}
	&& h'_\sigma(\vec r) = \frac{dv_H(\vec r)}{dN} + \frac{dv_{xc,\sigma}^{LDA}(\vec r)}{dN} \\
	\label{eq:g_new} 
	&& g'_\sigma(\vec r) = \frac{dv_{x,\sigma}^{EXX}(\vec r)}{dN} -  \frac{dv_{x,\sigma}^{LDA}(\vec r)}{dN}. 
\end{eqnarray}
%--------------
Note that $h_\sigma'$ is actually the same as $h_\sigma$ in Eq.~\ref{eq:h}. The performance of this new mixing scheme is tested on the molecular systems in Fig.~\ref{fig:IE}. For most systems, its performance is similar to the previous mixing scheme. However, for many anions, this new mixing scheme gives too high HOMO eigenvalues. This means that we also need to include $p_\sigma$ in the mixing parameter calculations if the orbital relaxation effect is considered in $h_\sigma$ and $g_\sigma$. Nevertheless, for H$_2$ and the CO molecule discussed in the next section. This new mixing scheme yields good predictions for the HOMO eigenvalues and is valid to use. For H$_2$, the HOMO eigenvalues predicted by the new and old schemes are $-15.87$ eV and $-16.02$ eV, respectively, which are close to the CCSD result of 16.39 eV. For CO, the HOMO eigenvalues from the new and old schemes are $-13.93$ eV and $-13.91$ eV, respectively, both of which agree well with the CCSD(T) result of 13.91 eV.

\begin{figure}[]
	\includegraphics[width=0.5\textwidth]{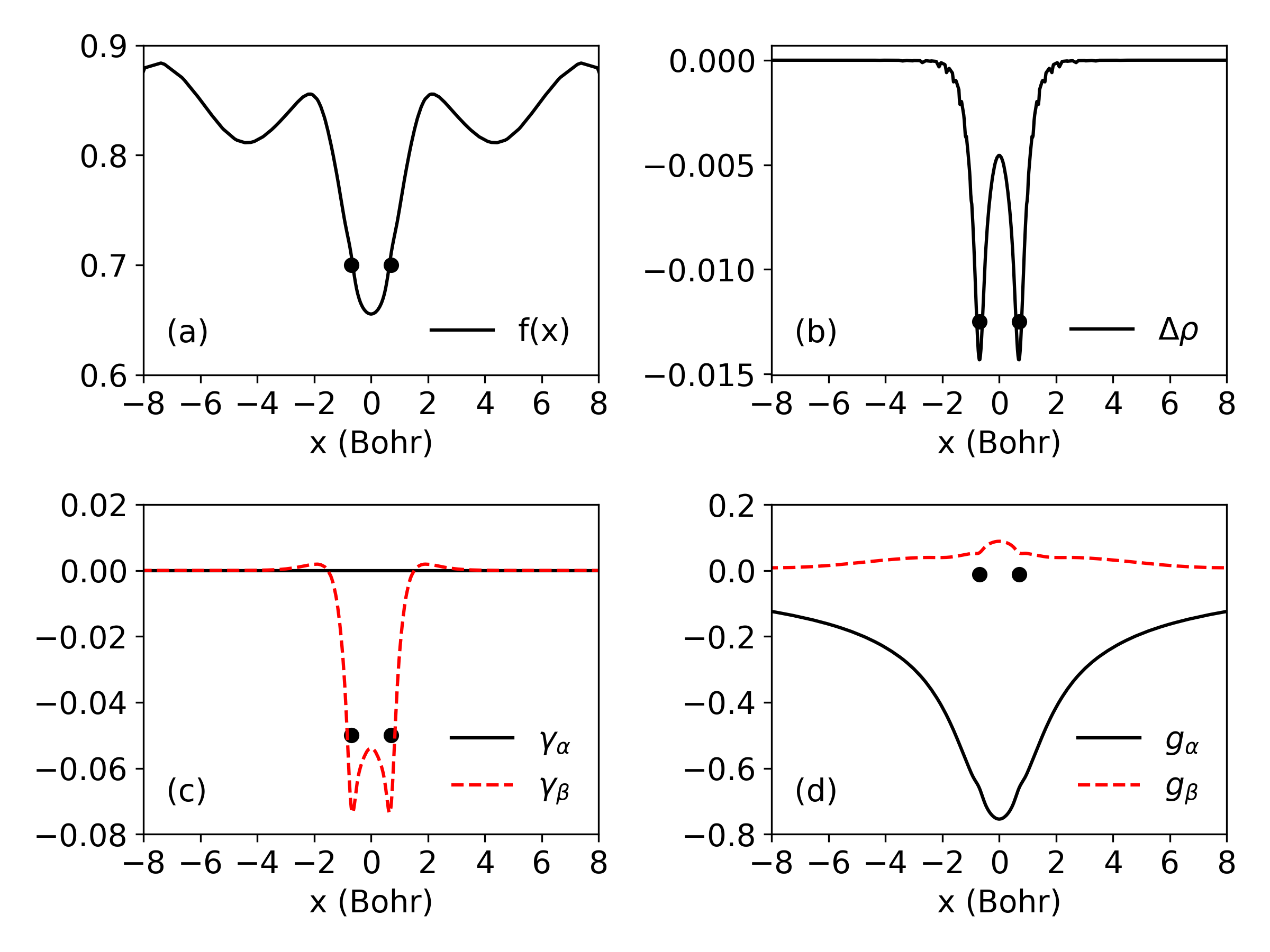}
	\caption{ Results for H$_2$ from the SMMP calculations in which $h_\sigma$ and $g_\sigma$ are calculated based on relaxed orbitals. The two H atoms are marked by the two black dots. (a) The mixing parameter. (b) The difference between the CCSD and SMMP densities. (c) $\gamma_\alpha$ and $\gamma_\beta$. (d) $g_\alpha$ and $g_\beta$. }
	\label{fig:H2_relax}
\end{figure}

The results based on the above new mixing scheme are summarized in Fig.~\ref{fig:H2_relax}. Even after including the orbital relaxation effect in $h_\sigma$ and $g_\sigma$, the mixing parameter is still less than one around the hydrogen atoms (Fig.~\ref{fig:H2_relax}(a)), leading to an incorrect positive $v_c$. 
This incorrect $v_c$ will cause errors in the electron density, which can be seen by comparing the SMMP and CCSD densities in Fig.~\ref{fig:H2_relax}(b). 
The density difference is defined as $\Delta \rho=\rho_\mathrm{SMMP} - \rho_\mathrm{CCSD}$. 
We see that $\rho_\mathrm{SMMP}<\rho_\mathrm{CCSD}$ around the hydrogen atoms, which confirms that the mixed XC potential around the atoms is too shallow. To have a deeper XC potential, the mixing parameter around the atoms needs to be increased. As discussed below, this can potentially be achieved by including $p_\sigma$ in the calculation.

To analyze the impact of $p_\sigma$ on the mixing parameter without explicitly calculating it, we examine $\gamma_\sigma$ which is related to $p_\sigma$ through Eq.~\ref{eq:p}. Since there is only one electron in the spin-$\alpha$ channel,  $\gamma_\alpha=0$. $\gamma_\beta<0$ in the bond region (especially around the hydrogen atoms) because the $\beta$ electron moves to the bond region to screen the hole created after removing the $\alpha$ electron. 
We can then estimate the sign of $p_\sigma$ based on Eq.~\ref{eq:p}. We find that $p_\alpha=0$ and $p_\beta>0$ around the two hydrogen atoms. 
We also need to know the signs of $g_\alpha$ and $g_\beta$, since the additional contribution to the mixing parameter is $\sum_\sigma p_\sigma g_\sigma/\sum_\sigma g_\sigma^2$ (see Eq.~\ref{eq:f}). 
Fig.~\ref{fig:H2_relax}(d) shows that $g_\alpha<0$ and $g_\beta>0$. Therefore, we have $p_\alpha g_\alpha=0$ and $p_\beta g_\beta>0$ in the bonding region.
This means that, if we include $p_\sigma$ in the calculations, the mixing parameter around the hydrogen atoms will increase, which then leads to a deeper XC potential.

\subsubsection{CO}

\begin{figure}[]
	\includegraphics[width=0.5\textwidth]{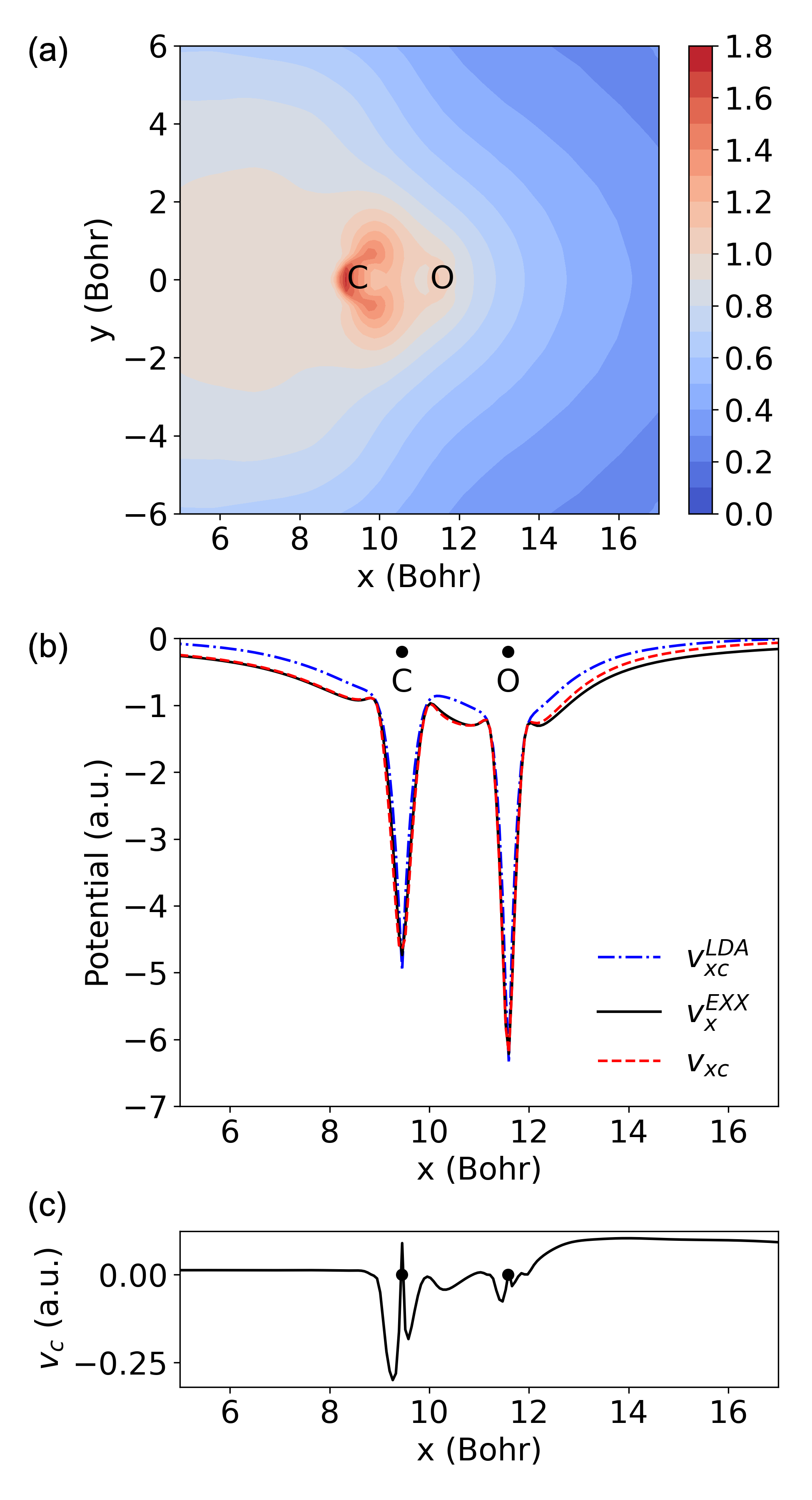}
	\caption{(a) Contour plot of the mixing parameter of CO from SMMP calculation using FOA. (b) Comparison of the XC potentials calculated using LDA, EXX, and SMMP (based on FOA).
	}
	\label{fig:CO}
\end{figure}

Next, we examine a more complicated case: CO molecule. The mixing parameter, the mixed XC potential, and the KS correlation potential are shown in Fig.~\ref{fig:CO}. 
The mixing parameter is close to one in the vacuum region on the carbon side, which is consistent with the asymptotic behavior given in Eq.~\ref{eq:fasym}. 
However, the mixing parameter is much less than one in the vacuum region on the oxygen side. We note that this does not contradicts the asymptotic behavior in Eq.~\ref{eq:fasym}.  Eq.~\ref{eq:fasym} is derived based on the assumption that HOMO dominates in the vacuum region. However, CO's HOMO (5$\sigma$) does not dominate on the oxygen side. Instead, CO's $4\sigma$ and $1\pi$ orbitals dominate on the oxygen side. Thus, the mixing parameter is less than one in the vacuum region near the oxygen. Nevertheless, we expect the mixing parameter to eventually approach to one at large $r$, since the HOMO decays more slowly than the other occupied orbitals.

In Fig.~\ref{fig:CO}(a), we find that the mixing parameter is larger than one in certain regions (for example, around the carbon and oxygen atoms). This contradicts the common belief that the mixing parameter should be within the range of 0 to 1. The reason for $f>1$ in these regions is for producing a negative $v_c$. This can be understood from Eq.~\ref{eq:vxc}. Because $v_{xc}^{LDA}>v_x^{EXX}$ in these regions, $f>1$ leads to $v_x<v_x^{EXX}$, which then gives a negative $v_c$.

\begin{figure}[]
	\centering
	\includegraphics[width=0.5\textwidth]{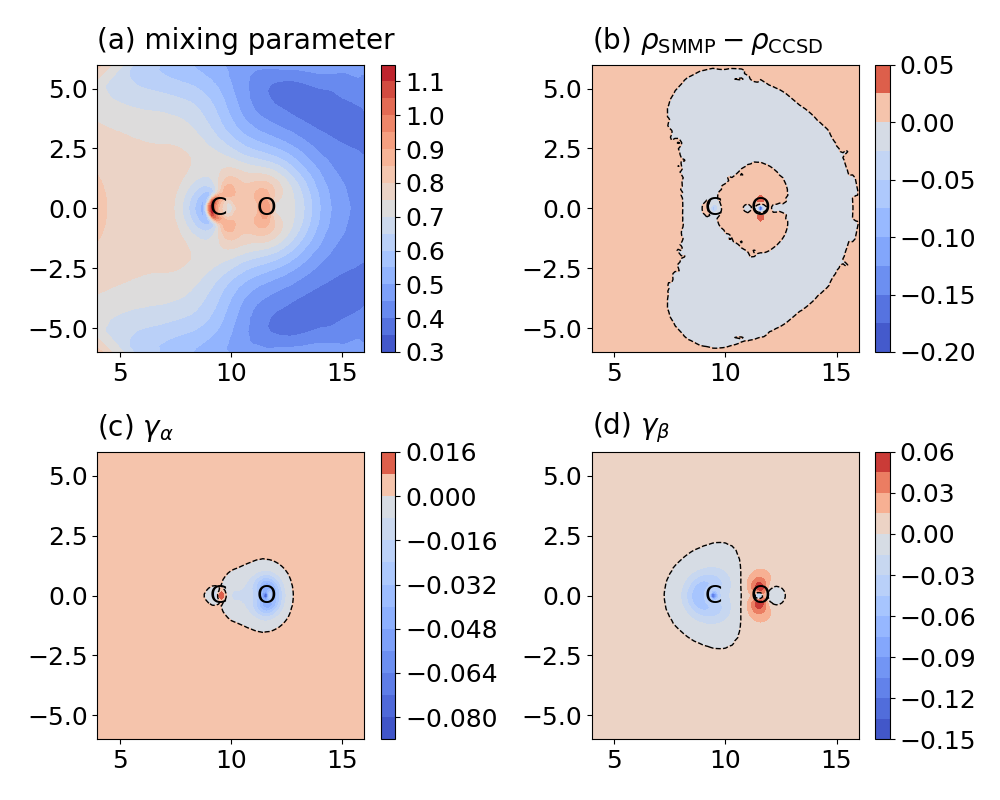}
	\caption{Results from SMMP calculations based on the mixing scheme Eq.~\ref{eq:new_mix}. 
		(a) The mixing parameter.
		(b) Density difference between SMMP and CCSD. 
		(c) $\gamma_\alpha$. (d) $\gamma_\beta$. 
		All results are in the atomic units. 
		In subplots (b), (c), and (d), the contour lines of zero are marked by dashed lines.}
	\label{fig:CO_relax}
\end{figure}

Next, we examine the impact of orbital relaxation on the mixing parameter. Similar to the case of H$_2$, we perform SMMP calculations using the new mixing scheme defined in Eq.~\ref{eq:new_mix}. $h_\sigma$ and $g_\sigma$ are evaluated based on the relaxed orbitals, and $p_\sigma$ is set to zero.
%-----------------
The difference between the SMMP and CCSD densities is given in Fig.~\ref{fig:CO_relax}(b).  $\rho_\mathrm{SMMP}>\rho_\mathrm{CCSD}$ around the oxygen and $\rho_\mathrm{SMMP}<\rho_\mathrm{CCSD}$ in most regions around the carbon. 
To examine whether the SMMP density can be improved by including $p_\sigma$ in the calculation, we calculate $\gamma_\alpha$ and $\gamma_\beta$ (Fig.~\ref{fig:CO_relax}(c) and (d)). In these calculations, the HOMO is from the spin $\alpha$. 
On the other hand, $g_\alpha<0$ and $g_\beta>0$ in the entire system.
%--------------- oxygen region ---------
Based on these results, we have $p_\alpha g_\alpha<0$ and $p_\beta g_\beta<0$ in the oxygen region, which will then reduce the mixing parameter there. This will produce a shallower XC potential around the oxygen atom, repelling the electrons towards the carbon atom. 
%--------------- carbon region ---------
With a similar analysis, including $p_\sigma$ in the calculation will increase the mixing parameter in the carbon region. This will lower the XC potential around the carbon atom, attracting more electrons to the carbon. The above analysis shows that including $p_\sigma$ in the mixing parameter calculation can correct the SMMP density as intended.

\section{Conclusions}

In this work, we developed a method, named spatial mixing of model potentials, aiming to construct reliable XC potentials by properly mixing the EXX and LDA potentials in real space. The mixing parameter is derived from the derivative discontinuity of electron density and is fully first-principle. 
%------------------
To avoid inverting the KS linear response during solving the mixing parameter, a frozen orbital approximation is introduced. Despite its simplicity, this approximation gives reasonable predictions for the ionization energies, fundamental gaps, and singlet-triplet gaps of various molecular systems. 
%--------------
We then investigated the impact of this approximation on the mixing parameters. The examples show that the XC potentials could be further improved by removing the approximation. This motivates a full calculation of the mixing parameter in future studies. 
%-------------------------
One limitation of our current formalism of SMMP is that it cannot be applied to systems having degenerate HOMOs. Future work will focus on resolving this issue.  
%-------------------------
Another limitation is that SMMP cannot be used for structure optimization because it lacks an XC energy functional and is not a variational method. One possible solution is to directly minimize forces by solving the equation $\vec F_{i}(\{\vec R_j\})=0$, where $\vec F_i$ is the force of atom $i$ and $\vec R_j$ is the coordinate of atom $j$. This set of equations can be solved using quasi-Newton methods, such as Broyden's method.\cite{broyden1965class} The forces can be calculated in the conventional way, by assuming that SMMP is a variational method.
The quality of the structure optimization is then determined by how well SMMP's results agree with the exact solutions. This will be explored in future work.
%------------
In summary, this work shows that the derivative discontinuity of electron density is a promising condition for determining the mixing of EXX and LDA potentials in real space. Further development in this direction could yield sufficiently accurate XC potentials for resolving electronic structures in many challenging systems, such as mixed-valence systems, radicals/biradicals, and polarons.

\section*{Supplementary material}
The supplementary material contains (a) the errors of the SMMP and EXX calculations for the molecular systems in Fig.~\ref{fig:IE}, and (b) the details of CCSD(T) energy calculations.

\begin{acknowledgments}
This work is supported by the National Science Foundation CAREER award (\#1752769). Some calculations were performed using the Extreme Science and Engineering Discovery Environment (XSEDE)\cite{Towns2014} supported by the National Science Foundation Grant No. ACI-1548562, with the allocation number CHE230067.
\end{acknowledgments}

\bibliography{ref}

\end{document}